# Ballistic quantum transport in L-shaped Vertical Halo-Implanted p$^+$-GaSb/InAs n-TFETs

Bhupesh Bishnoi and BahnimanGhosh

*Abstract*—In the present work, we have investigated ballistic quantum transport in vertical halo-implanted p$^+$-GaSb/InAs n-TFETs. We have investigated the current–voltage characteristics, ON-current, OFF-current leakage, subthreshold swing variation as function of gate length, drain length, gate undercut, equivalent oxide thickness, High-K and drain thickness. The electrostatic control, I-V performances and optimization of device structure are carried out for novel L-shaped nonlinear geometry n-TFETs. In the n-TFETs device p$^+$-GaSb/InAs heterostructure gives rise to type-III broken gap band alignment. In this geometry the gate electric field and tunnel junction internal field are oriented in same direction and assist the Band-to-Band tunnelling process. To study the ballistic quantum transport in this L-shaped nonlinear geometry we used 3-D, full-band atomistic $sp^3d^5s^*$spin-orbital coupled tight-binding method based quantum mechanical simulator which works on the basis of Non-Equilibrium Green Function formalism to solve coupled Poisson-Schrödinger equation self-consistently for potentials and Local Density of state.

## I. INTRODUCTION

Intensive research work is going on in TFETs as power-supply scaling below 0.5 V is possible in these devices and at low voltages TFETs can outperform aggressively scaled MOSFETs. Hence, overall power consumption can be reduced in nanoelectronics integrated circuits by using TFETs. [1]
In TFETs charge carriers are injected into the channel by band-to-band tunneling (BTBT) process and hence compared to conventional MOSFETs TFETs can have subthreshold swing lesser than 60 mV/decade. [2] In present scenario Tunnel field-effect transistors (TFETs) are promising candidates due to their steep subthreshold swing(SS), better ON to OFF current ratio and high drive current at low voltages operation.In the ITRS 2012 roadmap TFETs operate on $V_{DD}$ lesser than 0.5 V, $I_{ON}$ current of 100 milli-amperes, $I_{ON}/I_{OFF}>10^5$ and SS below 60 mV per decade.[2]Recently,

Bhupesh Bishnoi is with the Department of Electrical Engineering, Indian Institute of Technology, Kanpur, 208016, INDIA (e-mail: bbishnoi@iitk.ac.in).
Bahniman Ghosh is with the Department of Electrical Engineering, Indian Institute of Technology, Kanpur, 208016, INDIA and Microelectronics Research Center, 10100, Burnet Road, Bldg. 160, University of Texas at Austin, Austin, TX, 78712, USA (e-mail: bghosh@utexas.edu).

new vertical geometry TFETs has been experimentally demonstrated. In this geometry steep subthreshold swing (SS) can be achieved by in line field orientation of gate field and tunnel junction internal field. [3] GaSb/InAs broken-gap (BG) band aligned materials have zero band overlap and narrow band gap of 0.7266 eV and 0.354 eV and are favorable to build TFETs. The GaSb and InAs junction gives type-III broken bandgap alignment which assists the Band-to-Band tunneling (BTBT) process at the junction. [4] In this article, we demonstrate various electrostatic and geometrical considerations that influence the scaling and design of vertical halo-implanted p$^+$-GaSb/InAs n-TFETs of 4 nm thin channel structures with gate lengths of 20 nm.

## II. DEVICE STRUCTURE

The vertical halo-implanted p$^+$-GaSb/InAs n-TFETs have n-type InAs channel thickness $T_{InAs}$ of 4 nm with a doping density of $5\times10^{17}$ cm$^{-3}$ and 10 nm p-type GaSb source with doping density of $4\times10^{18}$ cm$^{-3}$. Figure 1 shows the cross section cartoon diagram of simulated vertical halo-implanted p$^+$-GaSb/InAs n-TFETs structures.

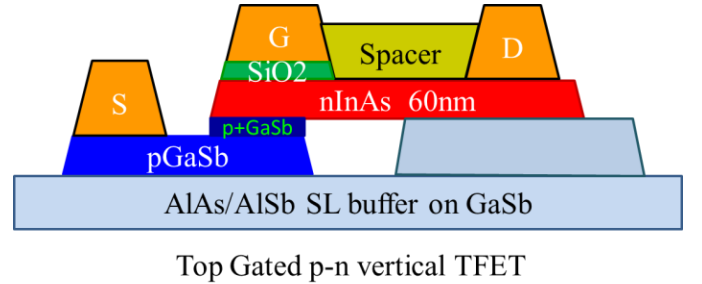

Fig.1. Cross section cartoon diagram of the simulated vertical halo-implanted p$^+$-GaSb/InAs n-TFETs

We have 2 nm halo-implanted p$^+$-type GaSb source injector in between source and drain region with doping density of $4\times10^{19}$ cm$^{-3}$. As per ITRS 2012 we performed simulation at 20 nm gate length ($L_G$), 10 nm source length, 60 nm drain length ($L_D$) and 1.9 nm SiO$_2$ gate thickness. SiO$_2$ spacer of an overlap length ($L_S$) is used to decouple the drain-gate region to reduce ambipolar conduction. Junction length ($L_J$) is length of active tunnelling junction and InAs channel has 10 nm undercut length ($L_{UC}$) which helps to achieve a steep SS. Figure 2 shows the structure of simulated device.

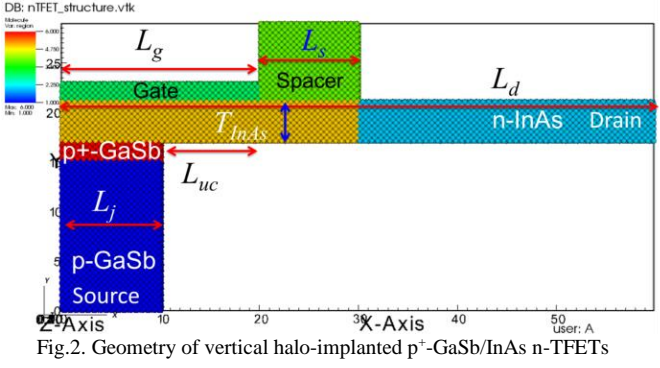

Fig.2. Geometry of vertical halo-implanted p$^+$-GaSb/InAs n-TFETs

## III. SIMULATION APPROACH

As vertical halo-implanted p$^+$-GaSb/InAs n-TFETs geometry is 2-D in nature and associated current path is in 2-D, analysis of these devices goes beyond the Wentzel-Kramers-Brillouin (WKB) approximation based 1-D tunnelling models in which confinement effects and band quantization effects are neglected. [5] The modification of density of states due to confinement is included in 3-D, full-band, quantum mechanical simulator based on atomistic $sp^3d^5s^*$ spin-orbital coupled tight-binding representation of the band structure which solves Schrödinger and Poisson equations self-consistently. [6-9] Carrier charge densities are self-consistently coupled to the calculation of electrostatic potential. In the simulator we also incorporate quantization effect due to narrow size and neglecting this will increase the band gap and underestimate band-to-band tunnelling probability. A full self-consistent quantum mechanical simulation including electron-phonon scattering can in principle describe TFETs with high accuracy. [10-13] But, such calculation needs extremely high computational resources to solve NEGF equations. [14] Hence, coherent transport is simulated to obtain upper device performance limit.[15] In the present work, a method combining Non-equilibrium Green's Function (NEGF) formalism with semi-classical density is used to achieve an efficient simulation of vertical halo-implanted p$^+$-GaSb/InAs n-TFETs. [16-21] Transport direction is along <100> crystal axis in the channel and surface orientation is along (100). In the active region of device every atom is represented by a matrix and in the simulation Schrödinger equation is solved for 8696 active atoms. Gate dielectric layer and spacer layer are modeled as imaginary materials layer which has infinite bandgap as they separate the gate contacts and InAs channel and do not participate in transport calculation. Hence, in the Poisson equation they are characterized by their relative dielectric constant.In the transport path 960 energy points and 31 momentum points are taken for calculation.

## IV. RESULTS

A. *Energy-position resolved local density of states LDOS (x, E) and energy-position resolved electron density spectrum $G_n$ (x, E)*

Figure 3 shows the energy-position resolved local density of states LDOS (x, E) and energy-position resolved electron density spectrum $G_n(x, E)$ of the vertical halo-implanted p$^+$-GaSb/InAs n-TFETs at $V_{DS}$= 0.3 V in the ON-state condition with variation of $V_{GS}$ from -0.1 V to 1.2 V in the step of 0.1 V. In the ON-state biasing condition, gate modulates the position of the channel barrier and channel conduction band is pulled down below the source valence band to increase the source injection as seen in Figure 3. The energy-position resolved electron density spectrum $G_n(x, E)$ as shown on log scale shows the occupation of LDOS (x, E) by the respective source and drain contact Fermi reservoirs at $V_{DS}$= 0.3 V.Figure 3 also shows the original heterojunction broken bandgap at the source-channel interface and band shift in BG characteristic due to quantization.

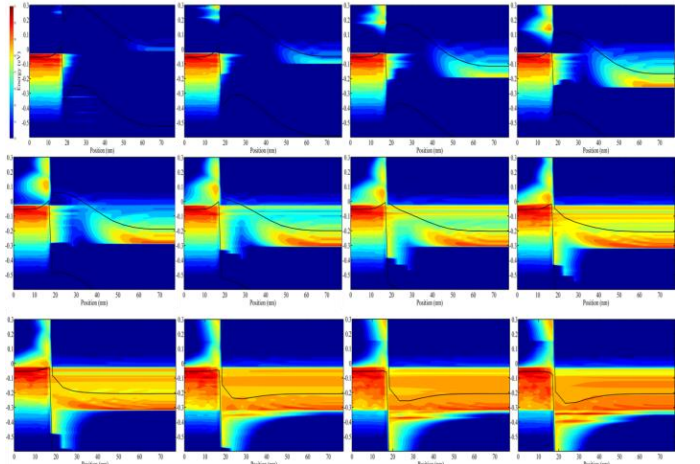

Fig.3. Energy-position resolved LDOS(x, E) and energy-position resolved electron density spectrum $G_n(x, E)$ in the ON-state of vertical halo-implanted p$^+$-GaSb/InAs n-TFETs

Figure 4 shows the energy-position resolved local density of states LDOS (x, E) and energy-position resolved electron density spectrum $G_n(x, E)$ of the vertical halo-implanted p$^+$-GaSb/InAs n-TFETs at $V_{DS}$= 0.03 V in the OFF-state condition with variation of $V_{GS}$ from -0.1 V to 1.2 V in the step of 0.1 V. OFF-state leakage current is mainly due to phonon absorption assisted tunnelling current. Figure 4 also shows the carrier thermalization due to phonon emissions in the drain region.

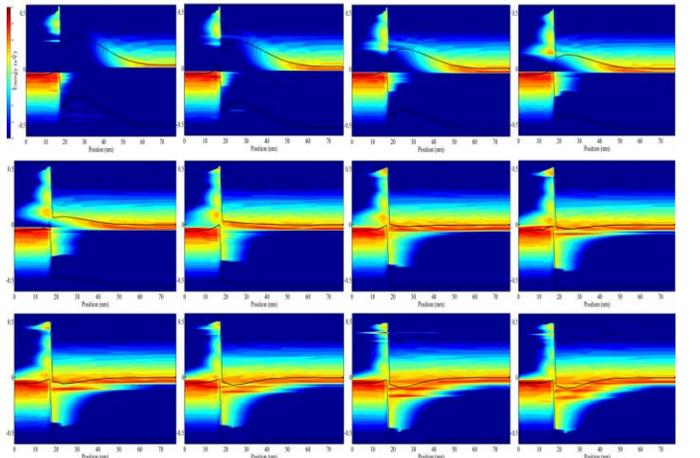

Fig.4. Energy-position resolved LDOS(x, E) and energy-position resolved electron density spectrum $G_n(x, E)$ in the OFF-state of vertical halo-implanted p$^+$-GaSb/InAs n-TFETs

## B. Current-Voltage characteristics

Figure 5 shows the $I_{DS}$-$V_{GS}$ transfer characteristics of vertical halo-implanted $p^+$-GaSb/InAs n-TFETs at $V_{DS}$ of 0.3V and 0.03V. In the ON-state condition with $V_{DS}$ of 0.3 V on applying gate voltage $V_{GS}$ of 0.6 V, an $I_{ON}$ current of 100 mA/μm, an $I_{ON}/I_{OFF}$ ratio of $10^{17}$ with subthreshold swing of about 15.34 mV/decade are obtained. In the ON-state for gate voltage $V_{GS}$ higher than 0.8 V drain current almost saturates and TFET gives high output resistance. In the OFF-state at $V_{DS}$= 0.3 V, minimum value of drain current is at $V_{GS}$= 0V, which indicates that gate has good control in turning off the tunnel junction.

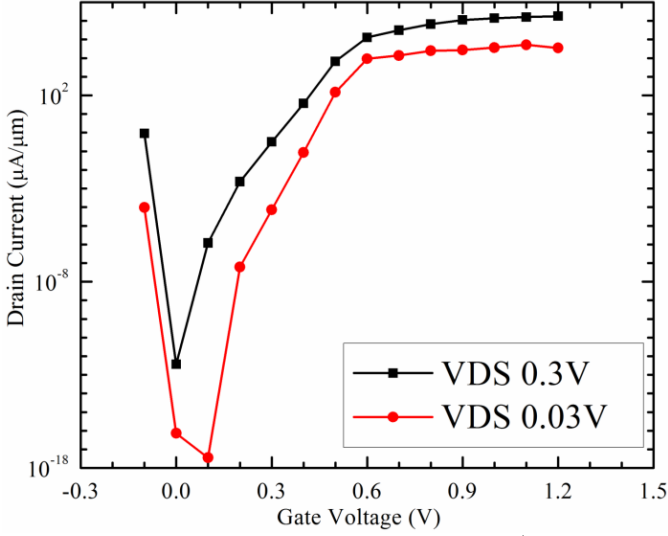

Fig.5 $I_{DS}$–$V_{GS}$ characteristics of the vertical halo-implanted $p^+$-GaSb/InAs n-TFETs at $V_{DS}$ = 0.03 V and $V_{DS}$ = 0.3 V.

Figure 6 shows the $I_{DS}$–$V_{DS}$ output characteristics of vertical halo-implanted $p^+$-GaSb/InAs n-TFETs at $V_{GS}$ variation of 0.1 V, 0.2 V, 0.3 V, 0.6 V, 0.9 V and 1.2 V. In the ON state, with drain voltage $V_{DS}$ of 0.3 V, the entire area of tunnelling junction is turned on and current density is uniform across the junction. For $V_{GS}$ of 0.1 V the tunnelling current is essentially extinguished and thermionic emission is governed by the leakage current. Figure 7 shows the ON-state $I_{DS}$–$V_{GS}$ transfer characteristics with variation of temperature in vertical halo-implanted $p^+$-GaSb/InAs n-TFETs. Temperature dependence is found to be considerably weaker in the temperature range of 200K to 300K clearly indicating direct band-to-band tunnelling. For gate voltage of 0.7 V and above, the temperature dependence is extremely minute and current saturates. At 400K we found that the current is an order of magnitude lower than that in the above mentioned temperature range in the gate voltage range of 0.2 V to 0.6 V. For gate voltage larger than 0.7 V current saturates and becomes independent of temperature.

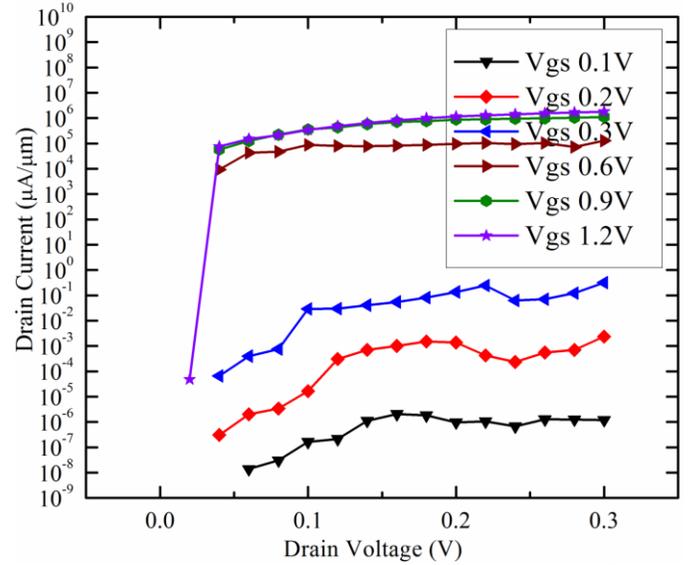

Fig.6. $I_{DS}$–$V_{DS}$ output characteristics of vertical halo-implanted $p^+$-GaSb/InAs n-TFETs with gate voltage variation $V_{GS}$

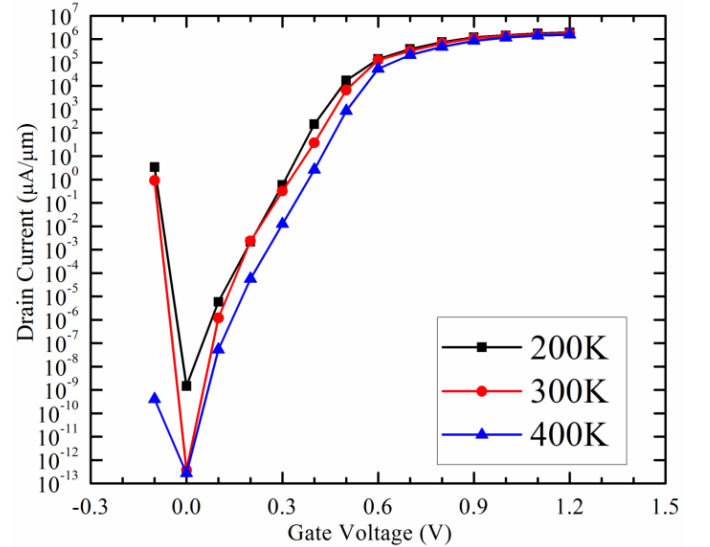

Fig.7. $I_{DS}$–$V_{GS}$ characteristics of the vertical halo-implanted $p^+$-GaSb/InAs n-TFETs at $V_{DS}$ = 0.3 V with variation in temperature.

## C. Effect of Geometry variation in vertical Halo-Implanted $p^+$-GaSb/InAs n-TFETs

We investigated five different types of geometric variations and their effect on I-V characteristics of vertical halo-implanted $p^+$-GaSb/InAs n-TFETs. Figure 8 shows the ON-state $I_{DS}$–$V_{GS}$ transfer characteristics with variation in drain length($L_D$) in the vertical halo-implanted $p^+$-GaSb/InAs n-TFETs. The ambipolar current is estimated in the real devices sense since the simulation includes the quantization effect and effective increase in channel bandgap. Electrostatic effect remains unaffected by the channel quantization and increase in channel bandgap. We observed that increasing the drain length ($L_D$) increases the ambipolar conduction current and hence turn-on characteristics and $I_{ON}$ current largely increases.

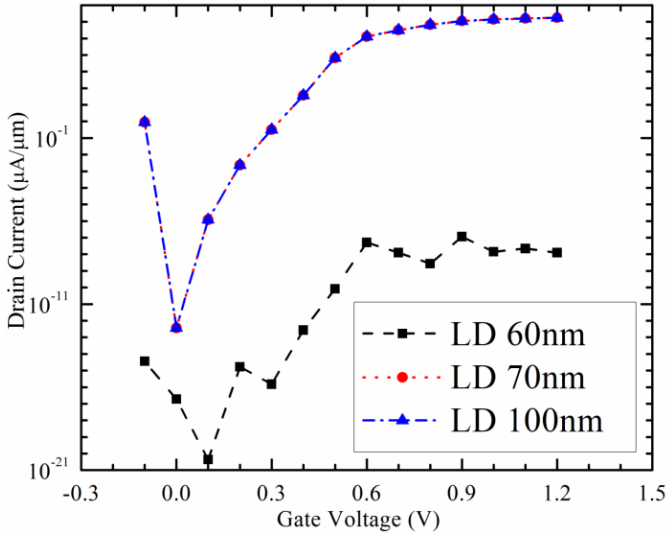

Fig.8. $I_{DS}$–$V_{GS}$ characteristics of the vertical halo-implanted $p^+$-GaSb/InAs n-TFETs at $V_{DS}$= 0.3 V with variation in drain length ($L_D$).

Figure 9 shows the ON-state $I_{DS}$–$V_{GS}$ transfer characteristics with variation in undercut lengths ($L_{UC}$) in the vertical halo-implanted $p^+$-GaSb/InAs n-TFETs. For zero undercut length, $I_{ON}/I_{OFF}$ ratio is reduced by ten orders of magnitude and the subthreshold swing increases to 43.4 mV/decade. Subthreshold swing reduces to 23.7 mV/decade on increasing undercut length to 5 nm and further subthreshold swing reduces to 13.8 mV/decade and $I_{ON}/I_{OFF}$ ratio increases on increasing the undercut length to 10 nm. $I_{ON}$ current reduces on further increase in the undercut length as $I_{ON}$ current is proportional to the tunnel junction area and hence on increasing the undercut length $L_{UC}$ width-normalized $I_{ON}$ current density decreases. Here the optimization criterion is overlapping the gate on the tunnelling region. The scalability of vertical halo-implanted $p^+$-GaSb/InAs n-TFETs is limited by undercut length $L_{UC}$ which is necessary to achieve a steep slope.

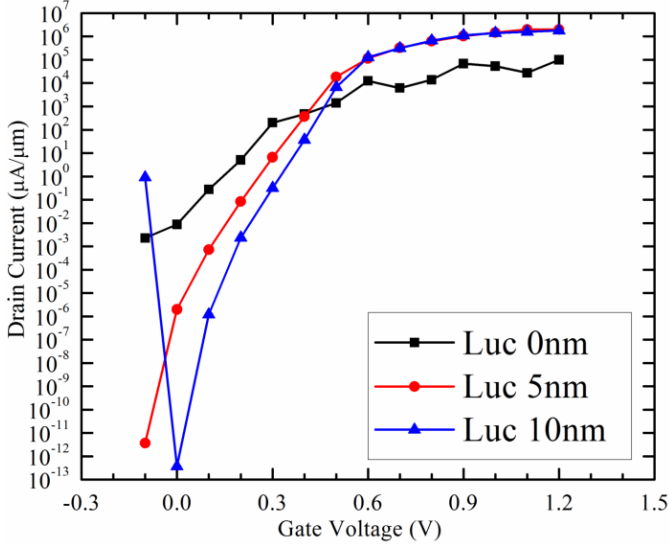

Fig.9. $I_{DS}$–$V_{GS}$ characteristics of the vertical halo-implanted $p^+$-GaSb/InAs n-TFETs at $V_{DS}$=0.3 V with variation in undercut length ($L_{UC}$).

In the figure 10 we showed the variation of equivalent oxide thickness ($T_{OX}$) on vertical halo-implanted $p^+$-GaSb/InAs n-TFETs. A thinner $T_{OX}$ gives strong coupling between InAs channel and gate field and hence gives rise to steeper subthreshold swing. With variation in $T_{OX}$ from 1.2 nm to 4 nm subthreshold swing increases from 13.8 mV/decade to 26 mV/decade. For thinner $T_{OX}$ due to strong coupling ambipolar current also increases and hence, $I_{ON}$ current slightly increases. For the vertical halo-implanted $p^+$-GaSb/InAs n-TFETs of given undercut lengths ($L_{UC}$) subthreshold slope depends upon the $T_{OX}$. In figure 11 we show the variation of different gate oxide material and its impact on the performance of vertical halo-implanted $p^+$-GaSb/InAs n-TFETs. For $Al_2O_3$ High-K gate material subthreshold swing reduces to 12 mV/decade and for $HfO_2$ High-K gate material subthreshold swing is 13 mV/decade.

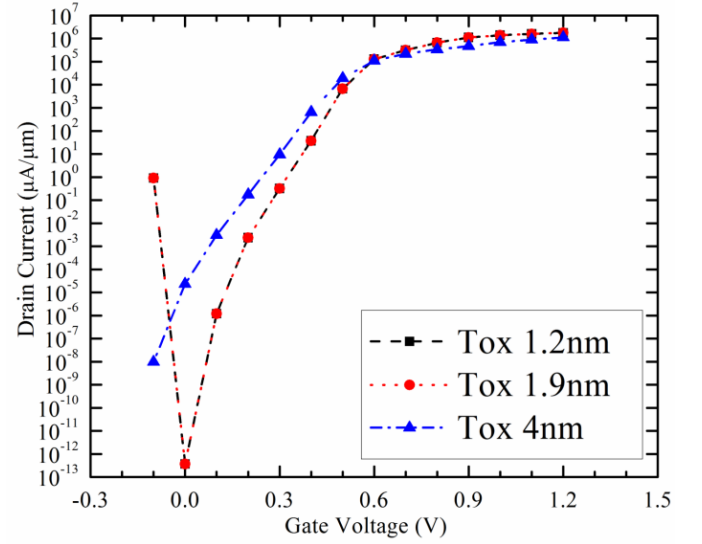

Fig.10. $I_{DS}$–$V_{GS}$ characteristics of the vertical halo-implanted $p^+$-GaSb/InAs n-TFETs at $V_{DS}$= 0.3 V with variation in Tox.

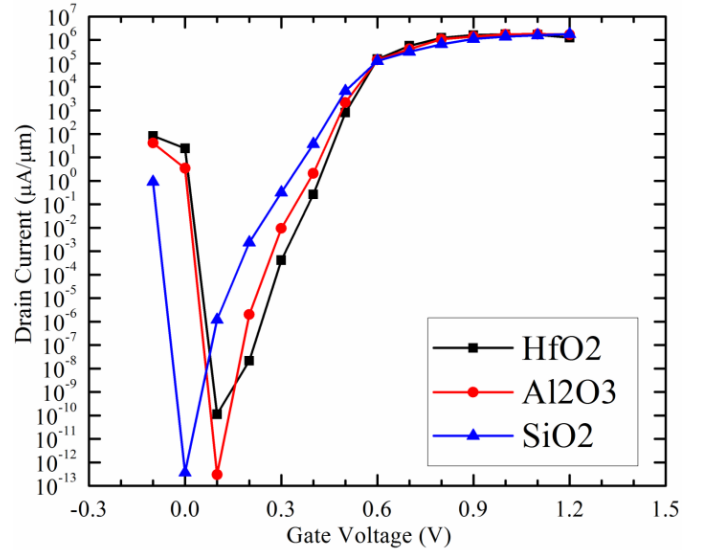

Fig.11. $I_{DS}$–$V_{GS}$ characteristics of the vertical halo-implanted $p^+$-GaSb/InAs n-TFETs at $V_{DS}$= 0.3 V with variation in gate oxide material.

In figure 12, we show the effect of gate-length ($L_G$)variation on vertical halo-implanted $p^+$-GaSb/InAs n-TFETs. We vary the gate lengths ($L_G$) from 10 nm to 30 nm while holding all other geometric parameters constant. Gate length ($L_G$) had a weaker influence on $I_{OFF}$ current. But, as gate length ($L_G$) increases, overlap over tunnelling junction area also increases and hence $I_{ON}$ current increases. $I_{ON}$ current increases with the gate length ($L_G$) from 10 nm to 30 nm. However, gate length strongly influences subthreshold swing and as gate length decreases subthreshold swing becomes steeper. Simulation results suggest that performance of vertical halo-implanted p+-GaSb/InAs n-TFETs is determined by drain-gate periphery and gate has good control at the center of the tunnel junction to turn off the device effectively.

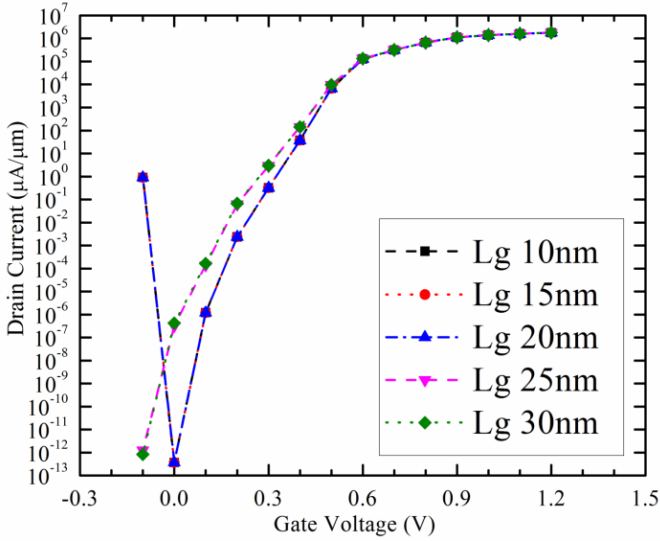

Fig.12. $I_{DS}$–$V_{GS}$ characteristics of the vertical halo-implanted p+-GaSb/InAs n-TFETs at $V_{DS}$= 0.3 V with variation in gate length ($L_G$)

Figure 13 shows the effect of variation of drain thickness ($T_{InAs}$) in vertical halo-implanted p+-GaSb/InAs n-TFETs. $I_{OFF}$ current and subthreshold slope depend upon the InAs channel thickness($T_{InAs}$).In the thicker InAs channel TFETs gate lose control on tunnel junction to shut off $I_{OFF}$ current effectively. We observe that at gate voltage of 0.6 V $I_{ON}$ current increases by more than two orders of magnitude when drain thickness ($T_{InAs}$) increases from 3 nm to 4 nm but with the overhead of higher value of subthreshold slope. For drain thickness ($T_{InAs}$) of 7nm, we observe that, at gate voltage of 0.5 V, $I_{ON}$ current of as high as 1 A/μm is achieved with $I_{OFF}$ of 0.1 μA/μm and subthreshold slope of 31 mV/decade. We also observe that high $I_{ON}$ current can be achieved by higher InAs concentration but at the cost of higher value of subthreshold slope.

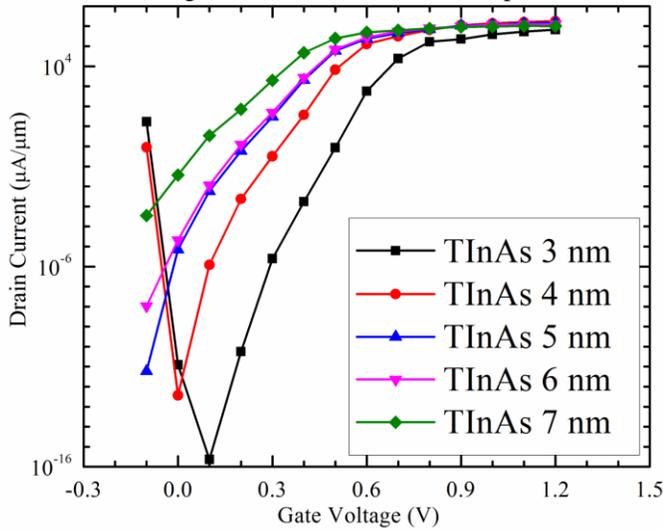

Fig.13. $I_{DS}$–$V_{GS}$ characteristics of the vertical halo-implanted p+-GaSb/InAs n-TFETs at $V_{DS}$= 0.3 V with variation in drain thickness ($T_{InAs}$).

Figure 14 shows ON-state $I_{DS}$–$V_{GS}$ characteristics of the vertical halo-implanted p+-GaSb/InAs n-TFETs with and without p+-GaSb halo implant. The effect of p+-GaSb halo-implant is clearly visible with increase in $I_{ON}$ current by 2 orders of magnitude at gate voltage of 0.6V. But, the SS slope increases from 11.85 mV/decade to 15.34 mV/decade.

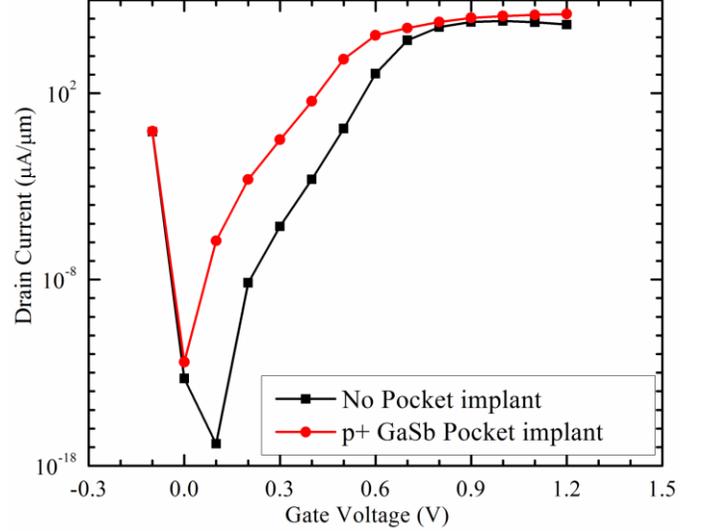

Fig.14. $I_{DS}$–$V_{GS}$ characteristics of the vertical halo-implanted p+-GaSb/InAs n-TFETs at $V_{DS}$= 0.3 V with and without p+-GaSb halo-implant.

## V. CONCLUSION

Vertical halo-implanted p+-GaSb/InAs n-TFETs of 4 nm thin channel structures with the gate lengths of 20 nm have been simulated using a 3-D, full-band, quantum mechanical simulator based on atomistic $sp^3d^5s^*$spin-orbital coupled tight-binding method. In the ON-state condition of $V_{DS}$ 0.3 V on applying gate voltage $V_{GS}$ of 0.6 V, an $I_{ON}$ current of 100 mA/μm, an $I_{ON}$/$I_{OFF}$ ratio of $10^{17}$ with subthreshold swing of about 15.34 mV/decade are obtained. We investigated five different types of geometric variations and their effects on I-V characteristics of vertical halo-implanted p+-GaSb/InAs n-TFETs. These simulation results suggest that the device performance of vertical halo-implanted p+-GaSb/InAs n-TFETs is determined by drain-gate periphery and gate length scaling rules do not work for TFETs. Instead other parameters such as channel thickness ($T_{InAs}$),undercut length ($L_{UC}$) and drain length ($L_D$) are critical design parameters for optimization of n-TFETs performance.Simulation results suggest that it is worthwhile to experimentally demonstrate and investigate vertical halo-implanted p+-GaSb/InAs n-TFETs in more detail as building block for future low power nano-electronics circuits.

## ACKNOWLEDGEMENT

The authors thank the Department of Science and Technology of the Government of India for partially funding this work.